\documentclass[12pt,letterpaper]{article}%
\usepackage{amsmath,amssymb,amsfonts}
\usepackage{color}
\usepackage{float}
\usepackage{hyperref}
\usepackage[Symbolsmallscale]{upgreek}
\usepackage{amsmath}
\usepackage{amsfonts}
\usepackage{amssymb,dsfont}
\usepackage{graphicx}
\usepackage{amssymb}%
\newcommand{\hhref}[1]{\href{http://arxiv.org/abs/#1}{arXiv:#1}}
\begin{document}
\begin{titlepage}
\begin{flushright}
LPTENS-10/35
\end{flushright}
\vskip 1.0cm
\begin{center}
{\Large \bf Central Charge Bounds in 4D Conformal Field Theory}
\vskip 1.0cm
{\large Riccardo Rattazzi$^{a}$,\ Slava Rychkov$^{b}$,\ Alessandro Vichi$^a$} \\[0.7cm]
{\it $^a$ Institut de Th\'eorie des Ph\'enom\`enes Physiques, EPFL,  CH--1015 Lausanne, Switzerland\\[5mm]
$^b$ Laboratoire de Physique Th\'{e}orique, \'{E}cole Normale Sup\'{e}rieure,\\
and Facult\'{e} de Physique, Universit\'{e} Pierre et Marie Curie,
France}
\end{center}
\vskip 2.0cm
\begin{abstract}
We derive model-independent lower bounds on the stress tensor
central charge $C_T$ in terms of the operator content of a
4-dimensional Conformal Field Theory. More precisely, $C_T$ is
bounded from below by a universal function of the dimensions of the
lowest and second-lowest scalars present in the CFT. The method uses
the crossing symmetry constraint of the 4-point function, analyzed
by means of the conformal block decomposition.
\end{abstract}
\vskip 2cm \hspace{0.7cm} September 2010
\end{titlepage}

\newpage

\section{Introduction}

Conformal Field Theory (CFT) was born to describe fixed points of
renormalization group flows. This still remains its main vocation, although it
has many other applications as well. In 2D, the constraining power of
conformal symmetry is tremendous and often leads to an exact solution of the
theory. In this paper we are concerned with the much less constrained 4D case.
Presumably, there are lots of interacting 4D CFTs out there, but we don't know
much about them. For instance, \textquotedblleft conformal windows" of gauge
theories should provide lots of examples, but even the spectrum of these
theories remains unknown (except for the chiral ring in the supersymmetric case).

In absence of an exact solution, it is natural to look for universal
constraints, satisfied all over the \textquotedblleft landscape" of CFTs. Two
years ago \cite{r1},\cite{r2} we found one such constraint, related to a gap
in the spectrum of operator dimensions. Namely, we examined the maximal
possible dimension of the lowest-dimension operator appearing in the Operator
Product Expansion (OPE) of two scalar operators. We found that if one fixes
the dimension of external scalars, the lowest-dimension operator that appears
cannot have a dimension above a certain model-independent bound.

Last year a different constraint was presented in \cite{cr}: it was found that
the OPE coefficient of three scalars cannot exceed a certain universal
$\mathcal{O}(1)$ bound, which depends only on their dimensions. One can call
this bound a universal limit on the interaction strength.

The above results were obtained by using consistency between OPE and crossing
symmetry (also known as OPE associativity) as a constraining principle. The
principle itself was first proposed more than 35 years ago by Polyakov
\cite{pol}, but until our work no general results were obtained from it.

We would like to mention a parallel line of development in 2D CFT, where
Hellerman \cite{Hellerman:2009bu} and others \cite{Gaberdiel:2008xb}%
,\cite{Hellerman:2010qd} have also studied constraints on the gap in the
spectrum, in terms of the central charge. Their main constraining principle is
modular invariance, which is limited to 2D, but morally not so different from
OPE associativity (both are related to the change of foliation when quantizing
the theory). Also interesting is the role played by all these universal
constraints in an ambitious program of exploring the space of CFTs initiated
by Douglas \cite{Douglas:2010ic}.

Coming back to 4D, in this note we will explore the following question:
\textit{What can we say about the central charge of the theory, if we know
something about the spectrum of its operator dimensions?} More precisely, we
will assume that the theory contains a scalar operator of a given dimension.
Under this assumption, we will show that the central charge must be bigger
that a certain universal lower bound. This is natural, since central charge
\textquotedblleft measures" the number of degrees of freedom, and by
assumption we know that our theory is not trivial.

In a certain range of scalar dimensions we will be able to show that the
central charge is necessarily bigger than that of the free scalar
(\textquotedblleft interacting theory has more degrees of
freedom\textquotedblright).

As we will explain below, the problem of bounding the central charge
\textit{from below }is equivalent to the problem of bounding \textit{from
above }the OPE coefficient of the stress tensor\ in the scalar times scalar
OPE. The similarity with the problem analyzed in \cite{cr} is then clear, and
we will be able to use the method of that paper. To make connection with the
results of \cite{r1},\cite{r2}, we will also study how the central charge
bound improves as a function of the assumed gap in the scalar sector of the OPE.

Everywhere we assume that we are dealing with a unitary theory. In a
non-unitary theory, central charge may well be zero (or negative) without CFT
being trivial, so our question would not even make sense.

\section{Formulation of the problem}

We begin by stating precisely our assumptions and goals.

In an arbitrary unitary CFT in $D=4$ spacetime dimensions, we consider the
central charge $C_{T}$, defined as the coefficient in the 2-point function of
the stress tensor operator $T_{\mu\nu}$:%
\begin{align}
\left\langle T_{\mu\nu}(x)T_{\lambda\sigma}(0)\right\rangle  &  =\frac{C_{T}%
}{S_{D}^{2}}\frac{1}{(x^{2})^{D}}\left[  \frac{1}{2}(I_{\mu\lambda}%
I_{\nu\sigma}+I_{\mu\sigma}I_{\nu\lambda})-\frac{1}{D}\delta_{\mu\nu}%
\delta_{\lambda\sigma}\right]  \,,\nonumber\\
I_{\mu\nu}  &  =\delta_{\mu\nu}-2x_{\mu}x_{\nu}/x^{2} \label{eq:TT}%
\end{align}
($S_{D}=2\pi^{D/2}/\Gamma(D/2)$). It is assumed that the stress tensor is
normalized canonically, that is consistently with the Ward identities (written
here schematically for scalars):%
\begin{equation}
\partial_{\mu}\left\langle T_{\mu\nu}(x)\phi(x_{1})\ldots\phi(x_{n}%
)\right\rangle =-\sum_{i}\delta(x-x_{i})\left\langle \phi(x_{1})\ldots
\partial_{\nu}\phi(x_{i})\ldots\phi(x_{n})\right\rangle \,. \label{eq:ward}%
\end{equation}
The central charge $C_{T}$ is an interesting quantity because it provides a
certain measure of the number of degrees of freedom in the theory. For
example, for a free conformal theory of $N_{\phi}$ scalars, $N_{\psi}$ Dirac
fermions, and $N_{A}$ vectors, we have \cite{op}%
\begin{equation}
C_{T}=\frac{4}{3}N_{\phi}+8N_{\psi}+16N_{A}. \label{eq:ctfree}%
\end{equation}
Moreover, by unitarity $C_{T}>0$, and $C_{T}=0$ corresponds to a trivial
theory. It is well known that $C_{T}$ is an imperfect measure since, unlike in
2D, it does not in general decrease along the RG flow \cite{capelli}. The
other central charge $a$, defined in terms of the 4D trace anomaly, fares
better in this respect \cite{a}, while still remaining imperfect \cite{Yuji}.
In this paper, we will stick to $C_{T}$ since it is the one which we are able
to constrain.

Now assume that our theory contains a primary Hermitean\footnote{If $\phi$ is
not Hermitean, we can consider its real and imaginary parts.} scalar operator
$\phi$ of a given dimension $d.$ Our main goal will be to show that under this
assumption, the central charge of the theory cannot become arbitrarily small.
In other words, there exists a certain universal bound%
\begin{equation}
C_{T}\geq f(d)>0, \label{eq:bound}%
\end{equation}
where $f(d)$ depends on $d$ but is otherwise model-independent. In this paper
we will derive such a bound in the interval $1\leq d\leq2.$

\section{Solution strategy}

\subsection{Conformal blocks}

We will approach this problem by imposing the constraint of OPE associativity
in the 4-point function of the operator $\phi$. Consider all primary Hermitean
operators appearing in the OPE $\phi\times\phi$:%
\begin{align}
\phi\times\phi &  \supset\mathds{1}+S_{\Delta}+\ldots(\text{spin
0)}\nonumber\\
&  +T_{\mu\nu}+\ldots\text{(spin 2)}\\
&  +\text{higher spins\thinspace.}\nonumber
\end{align}
Here in the first line we included the unit operator and all scalar primaries,
starting from a certain dimension $\Delta\geq1$ and higher. In the second line
we have the stress tensor (spin 2, dimension 4 primary) and possibly higher
dimension spin 2 fields. The third line contains all higher spin primaries
($\Delta\geq l+2$ by the unitarity bounds \cite{mack}). Note that by
permutation symmetry of the $\phi\phi$ state only even spins can appear in
this OPE.

Now, it has been shown by Dolan and Osborn \cite{do1} that every primary spin
$l,$ dimension $\Delta$ operator $O_{\Delta,l}$ appearing in the $\phi
\times\phi$ OPE with a coefficient $c_{\Delta,l}$ gives a contribution to the
4-point function of $\phi$ of the following form:%
\begin{gather}
\left\langle \phi(x_{1})\phi(x_{2})\phi(x_{3})\phi(x_{4})\right\rangle \supset
c_{\Delta,l}^{2}\frac{g_{\Delta,l}(u,v)}{(x_{12}^{2})^{d}(x_{34}^{2})^{d}%
},\nonumber\\
u=x_{12}^{2}x_{34}^{2}/(x_{13}^{2}x_{24}^{2}),\quad v=x_{14}^{2}x_{23}%
^{2}/(x_{13}^{2}x_{24}^{2})\,, \label{eq:contr}%
\end{gather}%
\begin{gather}
g_{\Delta,l}(u,v)=\frac{(-)^{l}}{2^{l}}\frac{k_{\Delta+l}(z)k_{\Delta
-l-2}(\bar{z})-(z\leftrightarrow\bar{z})}{z-\bar{z}}\,,\nonumber\\
k_{\beta}(x)\equiv x^{\beta/2+1}{}_{2}F_{1}\left(  \beta/2,\beta
/2,\beta;x\right)  \,,\label{eq:confblock}\\
u=z\bar{z},\quad v=(1-z)(1-\bar{z})\,.\nonumber
\end{gather}
The functions of the cross-ratios $g_{\Delta,l}(u,v)$ are called conformal
blocks. They can be thought of as summing up the contributions of the primary
$O_{\Delta,l}$ and all its descendants to the 4-point function of $\phi$, when
applying the OPE in the $(12)(34)$ channel. It is nontrivial that such
summation can be performed in closed form. The clear advantage of the
representation (\ref{eq:contr}) is that it can be used at finite point
separation, unlike the OPE which is only useful in the coincidence limit
$x_{i}\rightarrow x_{j}$.

\subsection{Normalizations}

Eq. (\ref{eq:contr}) assumes that both $\phi$ and $O$ are unit normalized:%
\begin{align}
\left\langle \phi(x)\phi(0)\right\rangle  &  =(x^{2})^{-d},\\
\left\langle O_{\mu_{1}\ldots\mu_{l}}(x)O_{\lambda_{1}\ldots\lambda_{l}%
}(0)\right\rangle  &  =\frac{1}{(x^{2})^{\Delta}}\left[  \frac{1}{l!}%
(I_{\mu_{1}\lambda_{1}}\ldots I_{\mu_{l}\lambda_{l}}+\text{perms}%
)-\text{traces}\right]  ,
\end{align}
and the coefficient $c_{\Delta,l}$ is extracted from the 3-point function%
\begin{align}
\left\langle \phi(x_{1})\phi(x_{2})O_{\mu_{1}\ldots\mu_{l}}(0)\right\rangle
&  =\frac{c_{\Delta,l}}{(x_{12}^{2})^{d-\frac{\Delta-l}{2}}(x_{1}^{2}%
)^{\frac{\Delta-l}{2}}(x_{2}^{2})^{\frac{\Delta-l}{2}}}\left(  Z_{\mu_{1}%
}\ldots Z_{\mu_{l}}-\text{traces}\right)  ,\label{eq:c1}\\
Z_{\mu}  &  =x_{1\mu}/x_{1}^{2}-x_{2\mu}/x_{2}^{2}.
\end{align}
At the same time for the stress tensor normalized canonically (\ref{eq:TT}%
),(\ref{eq:ward}), the 3-point function coefficient is fixed by the Ward
identity \cite{op}:%
\begin{equation}
\left\langle \phi(x_{1})\phi(x_{2})T_{\mu\nu}(0)\right\rangle =-\frac
{Dd}{(D-1)S_{D}}\frac{1}{(x_{12}^{2})^{d-1}x_{1}^{2}x_{2}^{2}}\left(  Z_{\mu
}Z_{\nu}-\frac{1}{D}\delta_{\mu\nu}Z^{2}\right)  \,.
\end{equation}
These relations determine the coefficient $c_{4,2}$ appearing in
(\ref{eq:contr}) and (\ref{eq:c1}) in terms of the central charge $C_{T}$ and
the dimension of $\phi$ \cite{do1}:%
\begin{equation}
c_{4,2}=-\frac{Dd}{D-1}\frac{1}{\sqrt{C_{T}}}\,. \label{eq:c42}%
\end{equation}

Via (\ref{eq:contr}), this crucial relation implies that for large $C_{T}$,
the contribution of the stress tensor to the 4-point function of $\phi$
decreases as $1/C_{T}$. This is in accord with what happens for example in
AdS/CFT \cite{adscft}, where $C_{T}\sim N^{2}$ while the stress tensor
contribution corresponds to the graviton exchange in the bulk, which is
$1/N^{2}$ suppressed. Theories without stress tensor (i.e. in which
$c_{4,2}=0$) should be viewed as theories with an infinite central charge. One
example of such a theory is the Gaussian scalar field of dimension $d>1$, see
\cite{penedones} for its conformal block decomposition.

However in this paper we are interested in constraining the opposite limit of
small $C_{T}$, in which the stress tensor contribution increases. As we will
see below, such an increase eventually becomes inconsistent with crossing
symmetry, and this will give us the bound (\ref{eq:bound}).

\subsection{Analytic structure of conformal blocks and crossing symmetry}

\begin{figure}[h]
\begin{center}
\includegraphics[
height=1.3908in,
width=2.6814in
]{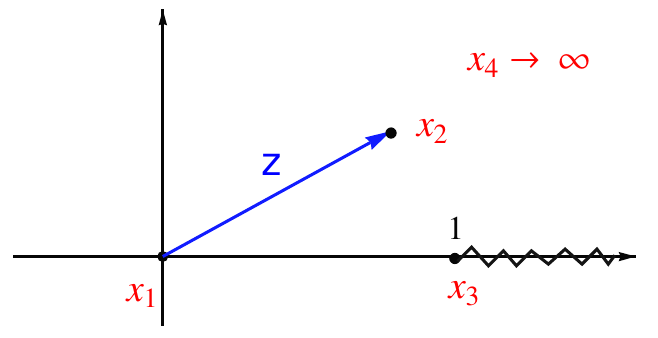}
\end{center}
\caption{\textit{Using conformal freedom, any configuration of 4 points can be
mapped into the one shown in this figure, in which 3 points are fixed and one
(}$x_{2}$\textit{) is moving in a two-plane passing through }$x_{1}$\textit{
and }$x_{3}$\textit{. The complex coordinate of }$x_{2}$\textit{ in this plane
is precisely the }$z$\textit{ of (\ref{eq:confblock}), while }$\bar{z}%
=z^{\ast}$\textit{. The conformal blocks are smooth everywhere in the plane
except for }$z=0$\textit{ and the shown }$(1,+\infty)$\textit{ cut along the
real axis.}}%
\label{fig:fig1}%
\end{figure}

Let us now discuss the analytic structure of the conformal blocks. The
variable $z$ appearing in (\ref{eq:confblock}) may look ad hoc, but in fact it
is the 4D analogue of the usual complex variable of the 2D CFT, see
Fig.~\ref{fig:fig1}. In the Euclidean signature $z$ is complex and $\bar
{z}=z^{\ast}$. In this case the conformal blocks are real functions, smooth
everywhere away from $z=0$ and from the $(1,+\infty)$ cut along the real axis.
Since the imaginary part of the hypergeometrics is discontinuous across the
cut, the conformal blocks have a $1/\operatorname{Im}z$ singularity there.
Everywhere else on the real axis, and away from it, they are
regular.

The asymptotic behavior of conformal blocks as $z\rightarrow0$ is fixed by the
OPE. The singular behavior in the $z\rightarrow1$ limit, which corresponds to
the crossed channel $x_{2}\rightarrow x_{3}$, has no simple physical meaning.
However, the sum over all blocks must be crossing symmetric and thus
consistent with the OPE in the crossed channel as well. Because of the
unphysicall singularities, it is not immediately clear how to impose the OPE
consistency in the crossed channel. In fact for this reason Polyakov
\cite{pol} suggested to use a different type of expansion, into objects he
called unitary blocks. However at the time the explicit and simple expressions
(\ref{eq:confblock}) were of course not yet known. Armed with these
expressions, a different strategy becomes possible.

Namely, we will study the crossing symmetry condition at finite point
separation, which can be written as%
\begin{equation}
\frac{G(u,v)}{u^{d}}=\frac{G(v,u)}{v^{d}}\,, \label{eq:crossing}%
\end{equation}
where we used the fact that crossing $x_{1}\leftrightarrow x_{3}$ corresponds
to the interchange of $u$ and $v$. Here $G(u,v)$ is the sum over all
contributing conformal blocks:%
\begin{equation}
G(u,v)=1+\sum c_{\Delta,l}^{2}\,g_{\Delta,l}(u,v)\,,
\end{equation}
where $c_{\Delta,l}^{2}$ are the squares of the OPE coefficients, and we
separated the contribution of the unit operator. It will be important that in
a unitary theory all $c_{\Delta,l}$ are real, so that their squares are
positive \cite{r1}. \begin{figure}[h]
\begin{center}
\includegraphics[
height=1.1841in,
width=1.3686in
]{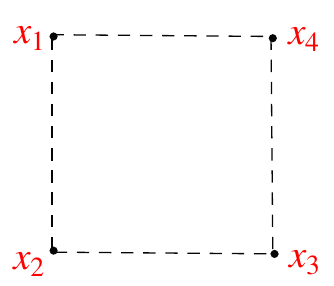}
\end{center}
\caption{\textit{This configuration, with 4 points at the vertices of a
square, is conformally equivalent to the one in Fig.~\ref{fig:fig1} with
}$z=1/2$\textit{.}}%
\label{fig:square}%
\end{figure}

Instead of going straight to the crossed OPE limit $x_{2}\rightarrow x_{3}$,
we will study (\ref{eq:crossing}) around the democratic
configuration\footnote{Similarly, Hellerman \cite{Hellerman:2009bu} in his
analysis of the modular invariance constraint chose to work at the selfdual
inverse temperature $\beta=2\pi$.} when $x_{2}$ is at equal distances from
$x_{1}$ and $x_{3}$. This corresponds to $z=1/2.$ In fact the same
configuration can be mapped conformally to 4 operators inserted at the
vertices of a square (Fig.~\ref{fig:square}). Both sides of (\ref{eq:crossing}%
) are regular around $z=1/2$. Expanding the crossing condition into a
two-dimensional power series around this point, we get a infinite number of
linear equations, which have to be satisfied for some positive coefficients
$c_{\Delta,l}^{2}$. Which $\Delta,l$ will enter the expansion with nonzero
coefficients depends on the CFT. The problem of unphysical singularities,
brought up by Polyakov \cite{pol}, is resolved as follows. The LHS of
(\ref{eq:crossing}) is smooth away from the cut along $(1,+\infty),$ while the
RHS away from ($-\infty,0)$, since the crossing maps $z\rightarrow1-z$.
Assuming that both sides can be analytically continued from their region of
convergence, the cuts \textit{must} cancel when summing over all $\Delta$ and
$l$. In other words, imposing that there is no cut should give no additional
constraints compared to the ones that we are using, although it may be a
different and perhaps a useful way to package the same information.

\subsection{Method of linear functionals}

For further discussion let us rewrite Eq.~(\ref{eq:crossing}) in the
equivalent `sum rule' form:%
\begin{align}
1  &  =\sum c_{\Delta,l}^{2}F_{d,\Delta,l}(u,v)\,,\label{eq:sumrule}\\
F_{d,\Delta,l}(u,v)  &  \equiv\frac{v^{d}g_{\Delta,l}(u,v)-u^{d}g_{\Delta
,l}(v,u)\,}{u^{d}-v^{d}}\,.
\end{align}
This equation says that the `crossing symmetry deficit'\ of all the fields in
the OPE, normalized to the deficit of the unit operator, has to sum up to $1$.

\begin{figure}[h]
\begin{center}
\includegraphics[]{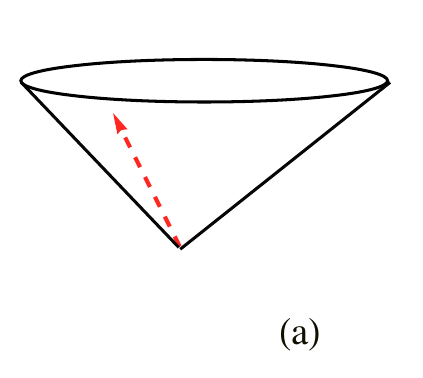}\hspace{1cm} \includegraphics[]{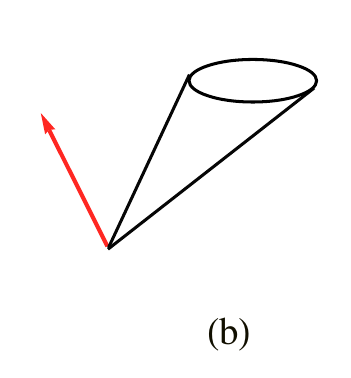}\hspace
{1cm} \includegraphics[]{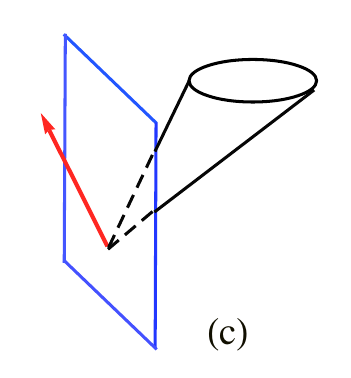}
\end{center}
\par
\label{fig:cone}\caption{\textit{Geometric interpretation of the sum rule: (a)
the sum rule has a solution $\Leftrightarrow$ $f\equiv1$ belongs to the cone;
(b) the assumed spectrum is such that the sum rule does not allow for a
solution $\Leftrightarrow$ $f\equiv1$ does not belong to the cone; (c) in the
latter situation, a hyperplane (the zero set of a linear functional) can be
found separating $f\equiv1$ from the cone.}}%
\end{figure}

Let us view Eq.~(\ref{eq:sumrule}) as a linear relation in the vector space of
functions of two variables $f(u,v)$. Then it can be given the following
geometric interpretation. As we keep the CFT spectrum fixed and vary the
squared OPE coefficients $c_{\Delta,l}^{2}\geq0$, the vectors in the RHS fill
in a \textit{convex cone} generated by the functions $F_{d,\Delta,l}$. The sum
rule says that the function $f(u,v)\equiv1$ must belong to this cone (see Fig
\ref{fig:cone}a).

If we start imposing restrictions on the CFT spectrum, for example by
demanding that there should be a gap in the scalar sector: $\Delta\geq
\Delta_{\ast}$ for $l=0$, this reduces the list of vectors generating the
cone, and \textit{a fortiori} the cone itself. It may well happen that the new
reduced cone no longer contains the function $f\equiv1$, Fig \ref{fig:cone}b.
A spectrum leading to such a cone cannot be realized in any CFT.

If the situation in Fig \ref{fig:cone}b occurs, then, since the cone is
convex, one can always find a hyperplane passing through the origin and
separating $f\equiv1$ from the cone, Fig \ref{fig:cone}c. In analytical
language, this means that there exists a linear functional $\Lambda$ taking
values of opposite sign on $f\equiv1$ and on the functions generating the
cone:%
\begin{equation}
\Lambda\lbrack1]\leq0,\quad\Lambda\lbrack F_{d,\Delta,l}]>0 \label{eq:lambda1}%
\end{equation}
In practice, the functional may be build up as a linear combination of the
partial derivatives with respect to $z$ and $\bar{z}$ at the democratic point
$z=\bar{z}=1/2$.\begin{figure}[h]
\begin{center}
\includegraphics[
height=1.1132in,
width=1.9851in
]{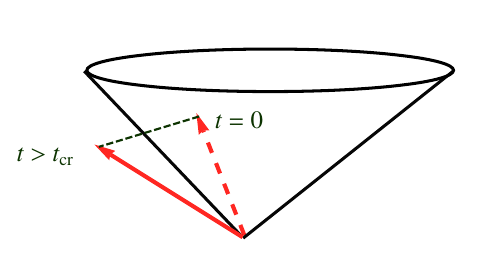}
\end{center}
\caption{\textit{Geometric interpretation of Eq.~(\ref{eq:sumrule1}). As }%
$t$\textit{ increases, the vector }$1-t\,F_{d,4,2}$\textit{ eventually leaves
the cone. }}%
\label{fig:moves}%
\end{figure}

So far we have described the method used in \cite{r1},\cite{r2} to constrain
the maximal allowed gap in the scalar sector. In order to constrain the size
of the OPE coefficient $c_{4,2}$, we proceed as follows \cite{cr}\footnote{We
choose $c_{4,2}$ for definiteness; the method in fact works for any OPE
coefficient.}. Let us rewrite the sum rule by transferring a part of the
stress tensor contribution into the LHS:%
\begin{equation}
1-t\,F_{d,4,2}=(c_{4,2}^{2}-t)F_{d,4,2}+\sum_{(\Delta,l)\neq(4,2)}c_{\Delta
,l}^{2}F_{d,\Delta,l} \label{eq:sumrule1}%
\end{equation}
The geometric interpretation of this equation is that the $t$-dependent vector
$1-t\,F_{d,4,2}(u,v)$ belongs to the same cone as before as long as $t\leq
c_{4,2}^{2}$. In other words, the maximal allowed value of $c_{4,2}^{2}$ can
be determined as the value $t=t_{cr}$ for which the curve $1-t\,F_{d,4,2}%
(u,v)$ crosses the cone boundary, Fig.~\ref{fig:moves}. Analytically, we can
detect that the crossing happened if there exists a linear functional such
that%
\begin{equation}
\Lambda\lbrack F_{d,\Delta,l}]\geq0 \label{eq:lambda2}%
\end{equation}
for all functions generating the cone, and
\begin{equation}
\Lambda\lbrack1-t\,F_{d,4,2}]=0. \label{eq:findt}%
\end{equation}
Note that in the present situation the function $f\equiv1$ must of course
belong to the cone, otherwise the CFT simply does not exist and there is no
point of discussing an upper bound on the OPE coefficients. Thus we are
assuming from the start $\Lambda\lbrack1]\geq0,$ unlike in (\ref{eq:lambda1}).

Since the functional is linear, Eq.~(\ref{eq:findt}) is satisfied for
\begin{equation}
t=\Lambda\lbrack1]/\Lambda\lbrack F_{d,4,2}],
\end{equation}
and for larger $t$ the functional will become negative as long as
$\Lambda\lbrack F_{d,4,2}]>0$. Thus we obtain the following result: each
functional $\Lambda$ satisfying (\ref{eq:lambda2}) gives a bound on the
maximal allowed value of $c_{4,2}^{2}$:%
\begin{equation}
\max c_{4,2}^{2}\leq\Lambda\lbrack1]/\Lambda\lbrack F_{d,4,2}]\,.
\label{eq:maxc42}%
\end{equation}
This bound can be optimized by choosing the functional judiciously.

The method just described was first applied in \cite{cr} to constrain the size
of the OPE coefficients of scalar operators, while here we will use it to
constrain the size of $c_{4,2}$, which via (\ref{eq:c42}) will give us a lower
bound on $C_{T}$. Another difference from \cite{cr} is that we will study how
the bound improves as a function of the assumed gap in the scalar sector of
the OPE.

\section{Results}

We will now present our numerical results. First of all, let us consider the
most general case when we are not making any assumption concerning the gap in
the scalar sector of the OPE. This means that the scalar operators appearing
in the OPE are allowed to have any dimension $\Delta\geq d.$ Operators with
lower dimensions are a priori excluded if $\phi$ is the lowest dimension
scalar. Under this assumption, we use the method of linear functionals to
bound $c_{4,2}^{2}$ from above. For this study, we choose linear functionals
of the form%
\begin{gather}
\Lambda\lbrack f]=\sum_{\substack{n,m\text{ even,}\\0\leq n+m\leq N}%
}\frac{\lambda_{n,m}}{n!m!}\partial_{a}^{n}\partial_{b}^{m}f|_{a=b=0}\\
z=1/2+a+b,\quad\bar{z}=1/2+a-b\,.
\end{gather}
As advertized, we are working around the democratic point $z=\bar{z}=1/2$. The
fact that we are choosing $z$ and $\bar{z}$ as real and independent can be
interpreted as a Wick rotation to the Minkowski space \cite{r1}. The
functional only contains even derivatives because the functions $F_{d,\Delta
,l}$ are even in both $a$ and $b$ \cite{r1}.

We will choose $\lambda_{0,0}=1$ to have $\Lambda\lbrack1]=1.$ Then to
optimize the bound (\ref{eq:maxc42}), the coefficients of the functional must
be chosen so that
\begin{equation}
\Lambda\lbrack F_{d,4,2}]\rightarrow\max\text{,} \label{eq:cost}%
\end{equation}
subject to the constraints (\ref{eq:lambda2}), which in our case mean%
\begin{align}
\Lambda\lbrack F_{d,\Delta,0}]  &  \geq0\text{\quad for all }\Delta\geq
d\,,\nonumber\\
\Lambda\lbrack F_{d,\Delta,l}]  &  \geq0\quad\text{for all }\Delta\geq
l+2,\quad l=2,4,\ldots\label{eq:constr}%
\end{align}

We will consider the functionals with the maximal derivative order up to
$N=16$.. Pushing to higher $N$ values is likely to somewhat improve the bound.
In principle $N$ as large as $18$ were demonstrated feasible in this kind of
studies \cite{r2}.

Eqs.~(\ref{eq:cost}),(\ref{eq:constr}) define an optimization problem for the
coefficients $\lambda_{m,n}$. The constraints are given by linear
inequalities, and the cost function is also linear, which makes it a
\textit{linear programming problem}. Although the number of constraints in
(\ref{eq:constr}) is formally infinite, they can be reduced to a finite number
by discretizing $\Delta$ and truncating at large $\Delta$ and $l$, where the
constraints approach a calculable asymptotic form \cite{r1}. The reduced
problem can be efficiently solved by well-known numerical methods, such as the
simplex method. A found solution can be then checked to see if it also solves
the full problem. This procedure was developed and successfully used in
\cite{r1},\cite{r2},\cite{cr}.

\begin{figure}[h]
\begin{center}
\includegraphics[
height=2.2591in, width=3.3829in ]{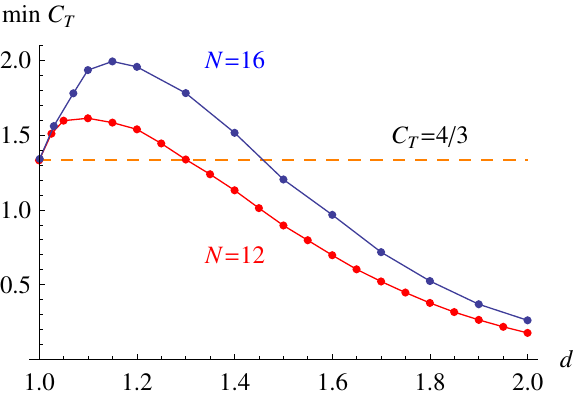}
\end{center}
\caption{\textit{The lower bound on the central charge $C_{T}$ in terms of the
dimension $d$ of the lowest-dimension scalar primary. The stronger bound
(upper blue curve) is obtained with $N=16$. For comparison we give a weaker
bound obtained with $N=12$ (lower red curve), which corresponds to the
horizontal axis $\Delta_{*}=d$ in the following Fig.\ \ref{fig:ct(d,delta)}.
The horizontal dashed line $C_{T}=4/3$ shows where our bound stays above the
free scalar central charge.}}%
\label{fig:ct(d)}%
\end{figure}

Using this procedure, we computed a bound on $c_{4,2}^{2}$ from above, which
via (\ref{eq:c42}) translates into a bound on $C_{T}$ from below. The latter
bound is plotted in Fig.~\ref{fig:ct(d)} as a function of the dimension of
$\phi$ in the range $1\leq d\leq2$. We plot our best bound for $N=16$ and, for
comparison, a weaker bound obtained with a smaller value $N=12$%
.

Postponing the discussion to the next Section, let us now consider what
happens with the bound in presence of a gap in the scalar spectrum. In other
words, we now assume that the first scalar operator in the $\phi\times\phi$
OPE has dimension $\Delta_{\ast}$ \textit{strictly bigger} than $d$.
Technically, this problem is analyzed exactly as the previous one, except that
the first set of constraints (\ref{eq:constr}) is replaced by a shorter list:%
\begin{equation}
\Lambda\lbrack F_{d,\Delta,0}]\geq0\text{\quad for all }\Delta\geq\Delta
_{\ast}\,.
\end{equation}
Because of considerable computer time involved, we solved this problem by
using linear functionals with $N=12$ only. The bound is given in
Fig.~\ref{fig:ct(d,delta)} as a contour plot in the $d,\Delta_{\ast}-d$ plane.
On the horizontal axis $\Delta_{\ast}=d$ the bound reduces to the $N=12$ bound
from Fig.~\ref{fig:ct(d)}. Naturally, when $\Delta_{\ast}$ increases, the
bound on $C_{T}$ gets stronger. The white region in upper left corresponds\ to%
\begin{equation}
\Delta_{\ast}>2+0.7(d-1)^{1/2}+2.1(d-1)+0.43(d-1)^{3/2} \label{eq:deltastar}%
\end{equation}
and is excluded, since such a large gap cannot be realized in any CFT
according to the results of \cite{r2}.

A text file with the coefficients of linear functionals used to derive the
shown bounds is included together with this arxiv submission.\begin{figure}[h]
\begin{center}
\includegraphics[
height=3.3829in, width=3.3829in
]{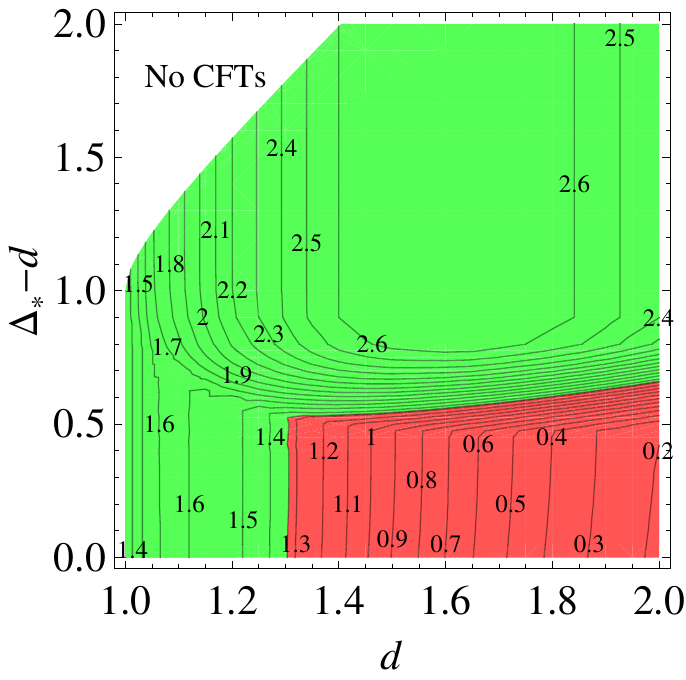}
\end{center}
\caption{\textit{Contour plot of the $C_{T}$ lower bound as a function of $d$
and of the gap $\Delta_{*}-d$, where $\Delta_{*}$ is the dimension of the
first scalar in the $\phi\times\phi$ OPE. The gap is nonnegative, since we
assume that $\phi$ is the lowest dimension scalar. On the horizontal axis the
bound reduces to the $N=12$ curve in Fig.\ \ref{fig:ct(d)}. The lighter green
color marks the region where the bound is above $C_{T}^{\mathrm{{free}}}=4/3$,
while in the darker red region the bound is below this value. As the gap
increases, the bound gets stronger, so that a rather weak assumption about the
gap is already enough to have $C_{T}>C_{T}^{\mathrm{{free}}}$.}}%
\label{fig:ct(d,delta)}%
\end{figure}

We end this section with a comment concerning the case of 2D conformal field
theories. Recall that in \cite{r1},\cite{r2} the maximal allowed gap in the
scalar spectrum was studied for the 2D case in parallel with 4D. This was
instructive since it allowed us to compare our bounds with the known OPEs in
the 2D minimal models. The analysis is feasible because the 2D conformal
blocks are known in a form just as simple as (\ref{eq:confblock}) (in odd
dimensions similarly simple expressions are not available). Analogously, in
the course of this project we have looked at the lower bounds on the Virasoro
central charge $c$ in the 2D CFTs, in the same $d,\Delta_{\ast}$ plane
($d\geq0$ as appropriate in the 2D case). We do not present them here because,
in the range that we considered, the found lower bounds were smaller than $1.$
Since all unitary 2D CFTs with $c<1$ are classified (these are precisely the
unitary minimal models \cite{friedan}), our bounds do not add any new
information in this case.

\section{Discussion}

Figs.~\ref{fig:ct(d)},\ref{fig:ct(d,delta)} contain our advertized main
results: universal lower bounds on the stress tensor central charge $C_{T}$.
Fig.~\ref{fig:ct(d)} gives a bound as a function of the dimension $d$ of the
lowest-dimension scalar $\phi$ present in the CFT. More precisely, the only
requirement on $\phi$ is that the OPE $\phi\times\phi$ not contain any scalar
of dimension less than $d$; this requirement is trivially satisfied if $\phi$
if lowest-dimension.

The first interesting point about this bound is that in the limit
$d\rightarrow1$ it approaches the free scalar central charge value
$C_{T}^{\text{free}}=4/3$, see Eq.~(\ref{eq:ctfree}). In other words, our
method shows that free theory limit is approached continuously. This is just
as in previous work, where we proved that as $d\rightarrow1$, the first scalar
in the $\phi\times\phi$ OPE must have dimension below $2$ \cite{r1},\cite{r2},
and the 3-point function $\left\langle \phi\phi\phi\right\rangle $ must
approach zero \cite{cr}.

Next, we see that for $1<d\lesssim1.4$ our bound stays \textit{above}
$C_{T}^{\text{free}},$ thus showing that an interacting theory necessarily has
larger central charge than the free one. This is also rather interesting.
Unfortunately, for larger $d$ our bound drops below $C_{T}^{\text{free}}$. We
do not know if this means that there are CFTs with $C_{T}<C_{T}^{\text{free}}%
$. More likely, this indicates that our bound is not best-possible in this
range. One could speculate that the best-possible bound should stay above
$C_{T}^{\text{free}}$ in the whole range $1<d<2$. The fact that it should
necessarily come down to $C_{T}^{\text{free}}$ (or lower) for $d=2$ can be
inferred by considering the dimension $2$ operator $\varphi^{2}$ in the free
scalar theory and its OPE with itself.

Note that we could also derive a bound without using the assumption
$\Delta\geq d$, which would be applicable to any scalar, not just the
lowest-dimension one. We have in fact derived also such a general bound,
although we do not show it here. We found that this general bound differs
little from the bound shown in Fig.\ \ref{fig:ct(d)} in the region of small
$d$, say for $d\lesssim1.3$. Thus this general bound could be useful if the
lowest dimension scalar has dimension very close to 1, while the second-lowest
is somewhat above 1. However, in the region of larger $d$, $d\gtrsim
1.7\div1.8$, the general bound drops to zero. This happens for the same
technical reason that the bounds on the scalar OPE coefficients in \cite{cr}
were blowing up around this value of the operator dimension. Because of this,
in this paper we focused on the lowest dimension scalar, which allowed us to
obtain a nontrivial bound in the full considered range of $d$.

Now let us discuss Fig.~\ref{fig:ct(d,delta)}, which gives the lower bound on
$C_{T}$ as a function of $d$ and $\Delta_{\ast}$. Here $\Delta_{\ast}$ is the
dimension of the lowest-dimension scalar in the OPE $\phi\times\phi$, assumed
to be above $d$. Again, this assumption is trivially satisfied if $\phi$ is
the lowest-dimension scalar present in the theory. Moreover, by the results of
\cite{r1},\cite{r2} $\Delta_{\ast}$ is limited from above by the bound given
in Eq.~(\ref{eq:deltastar}).

On the horizontal axis $\Delta_{\ast}=d$ the bound in
Fig.~\ref{fig:ct(d,delta)} reduces to the one shown in Fig.~\ref{fig:ct(d)},
while for larger $\Delta_{\ast}$ it naturally gets stronger. In fact we see
that $\Delta_{\ast}$ somewhat bigger than $d$ is already sufficient to raise
the bound above $C_{T}^{\text{free}}$ for all $d$ (the lighter green region in
the plot). The points with $\Delta_{\ast}\sim2d$ (i.e.~with an approximate
factorization of operator dimensions) belong to the green region by a big margin.

In summary, we have shown in this work that if a unitary 4D CFT is non-trivial
(in that it contains at least one primary scalar operator), then its central
charge $C_{T}$ cannot be arbitrarily low. We presented a universal bound on
$C_{T}$ as a function of the dimensions of the lowest and second-lowest
scalar. We hope that these bounds will be helpful in future efforts to chart
the \textquotedblleft landscape" of 4D conformal theories.

A relation like the one we derived, viewed from the AdS/CFT perspective
(although of course we cannot do it since we are not at large $N$), would
represent a lower bound on the Planck mass. One could then speculate that our
result belongs to the same class of constraints on quantum field theory as the
\textit{gravity as the weakest force} conjecture \cite{nic}. Unfortunately our
result cannot be directly applied to a phenomenon as fascinating and
unavoidabe as gravity, but it has a non-negligible consolation that it follows
from a rigorous mathematical analysis.

We believe more general constraints of the type discussed in this paper lie
ahead ready to be uncovered.

\textbf{Note added.} The morning of the day this paper was submitted to arxiv,
a nice paper \cite{poland} appeared which, among other things, also derives
lower bounds on the central charge.

\subsection*{Acknowledgements}

The work of R.R. and A.V. is supported by the Swiss National Science
Foundation under contract No. 200020-126941. The work of S.R. was supported in
part by the European Programme \textquotedblleft Unification in the LHC Era",
contract PITN-GA-2009-237920 (UNILHC).


\begin{thebibliography}{99}                                                                                               %


\bibitem {r1}{ R.~Rattazzi, V.~S.~Rychkov, E.~Tonni and A.~Vichi,
\textquotedblleft Bounding scalar operator dimensions in 4D
CFT,\textquotedblright\ JHEP \textbf{0812}, 031 (2008)
\href{http://arxiv.org/abs/0807.0004}{arXiv:0807.0004}.
}

\bibitem {r2}V.~S.~Rychkov and A.~Vichi, \textquotedblleft Universal
Constraints on Conformal Operator Dimensions,\textquotedblright%
\ Phys.\ Rev.\ D \textbf{80}, 045006 (2009)
\href{http://arxiv.org/abs/0905.2211}{arXiv:0905.2211}.


\bibitem {cr}F.~Caracciolo and V.~S.~Rychkov, \textquotedblleft Rigorous
Limits on the Interaction Strength in Quantum Field Theory,\textquotedblright%
\ Phys.\ Rev.\ D \textbf{81}, 085037 (2010)
\href{http://arxiv.org/abs/0912.2726}{arXiv:0912.2726}.


\bibitem {pol}A.~M.~Polyakov, ``Nonhamiltonian approach to conformal quantum
field theory,'' Zh.\ Eksp.\ Teor.\ Fiz.\ \textbf{66}, 23 (1974).


\bibitem {Hellerman:2009bu}S.~Hellerman, ``A Universal Inequality for CFT and
Quantum Gravity,'' \href{http://arxiv.org/abs/0902.2790}{arXiv:0902.2790}
[hep-th].


\bibitem {Gaberdiel:2008xb}M.~R.~Gaberdiel, S.~Gukov, C.~A.~Keller,
G.~W.~Moore and H.~Ooguri, ``Extremal N=(2,2) 2D Conformal Field Theories and
Constraints of Modularity,''
\href{http://arxiv.org/abs/0805.4216}{arXiv:0805.4216} [hep-th].


\bibitem {Hellerman:2010qd}S.~Hellerman and C.~Schmidt-Colinet,
\textquotedblleft Bounds for State Degeneracies in 2D Conformal Field
Theory,\textquotedblright%
\ \href{http://arxiv.org/abs/1007.0756}{arXiv:1007.0756} [hep-th].


\bibitem {Douglas:2010ic}M.~R.~Douglas, \textquotedblleft Spaces of Quantum
Field Theories,\textquotedblright%
\ \href{http://arxiv.org/abs/1005.2779}{arXiv:1005.2779} [hep-th].


\bibitem {op}H.~Osborn and A.~C.~Petkou, \textquotedblleft Implications of
Conformal Invariance in Field Theories for General
Dimensions,\textquotedblright\ Annals Phys.\ \textbf{231}, 311 (1994)
\href{http://arxiv.org/abs/hep-th/9307010}{arXiv:hep-th/9307010}.


\bibitem {capelli}A.~Cappelli, D.~Friedan and J.~I.~Latorre, ``C Theorem And
Spectral Representation,'' Nucl.\ Phys.\ B \textbf{352}, 616 (1991).


\bibitem {a}J.~L.~Cardy, ``Is There a c Theorem in Four-Dimensions?,''
Phys.\ Lett.\ B \textbf{215}, 749 (1988);
I.~Jack and H.~Osborn, ``Analogs For The C Theorem For Four-Dimensional
Renormalizable Field Theories,'' Nucl.\ Phys.\ B \textbf{343}, 647 (1990);
D.~Anselmi, J.~Erlich, D.~Z.~Freedman and A.~A.~Johansen, ``Positivity
constraints on anomalies in supersymmetric gauge theories,'' Phys.\ Rev.\ D
\textbf{57}, 7570 (1998)
\href{http://arxiv.org/abs/hep-th/9711035}{arXiv:hep-th/9711035};
K.~A.~Intriligator and B.~Wecht, ``The exact superconformal R-symmetry
maximizes a,'' Nucl.\ Phys.\ B \textbf{667}, 183 (2003)
\href{http://arxiv.org/abs/hep-th/0304128}{arXiv:hep-th/0304128}.


\bibitem {Yuji}A.~D.~Shapere and Y.~Tachikawa, \textquotedblleft A
counterexample to the 'a-theorem',\textquotedblright\ JHEP \textbf{0812}, 020
(2008) \href{http://arxiv.org/abs/0809.3238}{arXiv:0809.3238} [hep-th].


\bibitem {mack}S.~Ferrara, R.~Gatto and A.~F.~Grillo, \textquotedblleft
Positivity Restrictions On Anomalous Dimensions,\textquotedblright%
\ Phys.\ Rev.\ D \textbf{9}, 3564 (1974);
{G.~Mack, \textquotedblleft All Unitary Ray Representations Of The Conformal
Group SU(2,2) With Positive Energy,\textquotedblright%
\ \href{http://projecteuclid.org/DPubS?service=UI&version=1.0&verb=Display&handle=euclid.cmp/1103900926}{Commun.\ Math.\ Phys.\ \textbf{55}%
, 1 (1977)}.
}

\bibitem {do1}{ F.~A.~Dolan and H.~Osborn, \textquotedblleft Conformal four
point functions and the operator product expansion,\textquotedblright%
\ Nucl.\ Phys.\ B \textbf{599}, 459 (2001)
\href{http://arxiv.org/abs/hep-th/0011040}{arXiv:hep-th/0011040}.
\textquotedblleft Conformal partial waves and the operator product
expansion,\textquotedblright\ Nucl.\ Phys.\ B \textbf{678}, 491 (2004)
\href{http://arxiv.org/abs/hep-th/0309180}{arXiv:hep-th/0309180}.
}

\bibitem {adscft}O.~Aharony, S.~S.~Gubser, J.~M.~Maldacena, H.~Ooguri and
Y.~Oz, \textquotedblleft Large N field theories, string theory and
gravity,\textquotedblright\ Phys.\ Rept.\ \textbf{323}, 183 (2000)
\href{http://arxiv.org/abs/hep-th/9905111}{arXiv:hep-th/9905111}.


\bibitem {penedones}I.~Heemskerk, J.~Penedones, J.~Polchinski and J.~Sully,
``Holography from Conformal Field Theory,'' JHEP \textbf{0910}, 079 (2009)
\href{http://arxiv.org/abs/0907.0151}{arXiv:0907.0151} [hep-th].


\bibitem {friedan}D.~Friedan, Z.~a.~Qiu and S.~H.~Shenker, \textquotedblleft
Conformal Invariance, Unitarity And Two-Dimensional Critical
Exponents,\textquotedblright\ Phys.\ Rev.\ Lett.\ \textbf{52}, 1575 (1984).


\bibitem {nic}N.~Arkani-Hamed, L.~Motl, A.~Nicolis and C.~Vafa,
``The string landscape, black holes and gravity as the weakest force,''
JHEP {\bf 0706}, 060 (2007)
\hhref{hep-th/0601001}.


\bibitem {poland}D.~Poland and D.~Simmons-Duffin, ``Bounds on 4D Conformal and
Superconformal Field Theories,''
\href{http://arxiv.org/abs/1009.2087}{arXiv:1009.2087} [hep-th].

\end{thebibliography}
\end{document}